\documentclass[pra,aps,preprint]{revtex4-1}
\usepackage{graphicx}
\usepackage{amsmath,amssymb}
\usepackage{dsfont}

\newcommand{\sts}{\vert_{t\rightarrow\infty}}
\newcommand{\sto}{\vert_{t\rightarrow 0}}
\newcommand{\Hq}{\mathcal{H}}

\newcommand{\hs}{\mathcal{H}_{\text{\tiny S}}}
\newcommand{\hl}{\mathcal{H}_{\text{\tiny L}}}
\newcommand{\hsl}{\mathcal{H}_{\text{\tiny SL}}}
\newcommand{\wsl}{\omega_{\text{\tiny SL}}}

\newcommand{\nn}{\nonumber}

\newcommand{\wo}{\omega_{\circ}}

\newcommand{\tr}{\text{Tr}}
\newcommand{\rh}{\rho_{s}}
\newcommand{\rl}{\rho_{L}^{eq}}

\newcommand{\ic}{\vert_{t=0}}
\newcommand{\etal}{and others }
\begin{document}
\title{Prethermalization in an open quantum system coupled to a spatially correlated Bosonic bath}
\author{Saptarshi Saha}
\email{ss17rs021@iiserkol.ac.in}
\author{Rangeet Bhattacharyya}
\email{rangeet@iiserkol.ac.in}
\affiliation{Department of Physical Sciences, Indian Institute of Science Education and Research Kolkata,\\
Mohanpur -- 741246, WB, India}

\begin{abstract}

A nearly-integrable isolated quantum many-body system reaches a quasi-stationary prethermal state before
a late thermalization. Here, we revisit a particular example in the settings of an open quantum
system. We consider a collection of non-interacting atoms coupled to a spatially correlated bosonic bath
characterized by a bath correlation length. Our result implies that the
integrability of the system depends on such a correlation length. If this length is much larger than the distance between the atoms, such a system behaves as a nearly integrable open quantum system.
We study the properties of the emerging prethermal state for this case, i.e., the state's lifetime, the
extensive numbers of existing quasi-conserved quantities, the emergence of the generalized Gibbs state, and
the scaling of von Neumann entropy, etc. We find that for the prethermal state, the maximum growth of
entropy is logarithmic with the number of atoms, whereas such growth is linear for the final steady state, which is
the Gibbs state in this case. Finally, we discuss how such prethermal states can have significant applications in quantum
entanglement storage devices. 

\end{abstract}

\maketitle

\section{Introduction}
\label{intro}
Thermalization of a many-body quantum system has been an intense area of research in the last few decades \cite{rigol2008, nandkishore2015}. Although the statistical properties of such systems at thermal equilibrium are well known since the pioneer works by Boltzmann and von Neumann \cite{Kardar2007,von_neumann_proof2010}, it is only recently that the nonequilibrium dynamics of such systems towards the equilibrium are being explored. Schr{\"o}dinger equation can only predict the microscopic unitary dynamics of such systems. Therefore,  an irreversible dynamical equation of motion is required to describe the thermalization process.

The path to equilibrium for many-body quantum systems can be understood using two approaches. First, for isolated many-body systems, the system will thermalize due to the interactions of its constituents. In such cases, it is assumed that a large part of the system acts as an environment of the smaller subsystem, which helps the subsystem to thermalize \cite{nandkishore2015}. In the last few years, ultracold atoms have provided a promising experimental platform to demonstrate the thermalization of an isolated quantum system \cite{Bloch2008, Polkonikov2011}. Second, in an open quantum system, a small subsystem is assumed to be connected to the external environment. For Markovian systems, it is further assumed that the environment has large degrees of freedom and is in equilibrium at a temperature. The interaction between the subsystem and the environment leads to the non-unitary dynamics of the subsystem towards the final steady state \cite{Breuer_2002, atom-photon}. Nuclear Magnetic Resonance (NMR) spectroscopy, Nitrogen vacancy (NV) center, cavity optomechanical systems, etc., are used in experiments to observe the nonequilibrium dynamics of such systems \cite{Rotter2015}.

It is well known that isolated many-body systems (i.e., integrable systems) fail to thermalize and reach a stationary state whose properties differ entirely from the thermal state \cite{langen2015,kinoshita2006}. For such integrable systems, a large number of independent observables are conserved as they commute with the system Hamiltonian \cite{Langen2016, kollar2011}. In general, due to the presence of conserved quantities, the steady-state configuration of such systems is given by the generalized Gibbs ensemble \cite{Langen2016,kollar2011,rigol2007}. On the other hand, in the absence of such conserved quantities, the system directly thermalizes for the non-integrable systems.  

An intriguing feature is observed when a non-commuting perturbation Hamiltonian is present along with the integrable Hamiltonian. In such cases, the system behaves as a nearly integrable system \cite{Langen2016, kollar2011}. There exists more than one timescale in the system. In the transient phase, the effect of the perturbation is negligible, so the quasi-conserved quantities constrain the dynamics of the system, and the system is trapped in the quasi-stationary state. After a sufficiently long time, the system thermalizes, as the quasi-conservation laws are broken due to the presence of the perturbation. The lifetime of the quasi-stationary state depends on the strength of the perturbation \cite{kollar2011}. The emergence of such a quasi-stationary state far from the thermal equilibrium is also known as prethermalization \cite{ueda_2020}. A prethermal state is defined as a transient state, where the initial memory of the system is partially kept due to the existence of the quasi-conserved quantities, in contrast, such initial memory is completely lost in the thermal state. Throughout the years, experiments have been performed to understand the behavior of the prethermal states that emerge in isolated many-body systems \cite{gring2012,langen2015, Adu2013}. In the presence of the periodic drive in isolated systems, people also observed the existence of such prethermal plateau, which is known the Floquet prethermalization \cite{kuwahara_floquetmagnus_2016, bukov_prethermal_2015, holthaus_floquet_2016, rubio-abadal_floquet_2020,beatrez_floquet_2021, peng_floquet_2021}. In Floquet systems, the presence of such prethermal states leads to several exotic phases like discrete time crystals (DTC) \cite{sacha_time_2018,beatrez_critical_2023}. 

For a Markovian-type open quantum system (OQS), in the absence of other interactions, the steady state is always a Gibbs state, where the thermal bath defines the temperature. Such a system's dynamical equation can be obtained using a coarse-grained Born-Markov quantum master equation (QME) \cite{Breuer_2002, atom-photon}. Such a quantum dynamical map is trace-preserving and completely positive \cite{manzano_short_2020}. The generator of the dynamical map is defined as Liouvillian $(\mathcal{\hat{L}})$ (the dynamical equation is $\frac{d \hat{\rh}}{dt} = \mathcal{\hat{L}} \hat{\rh}$). The Liouvillian is essentially a non-Hermitian matrix, with at least a single zero eigenvalue; here, the zero eigenvalue appears due to the trace preservation \cite{albert_symmetries_2014}. The right eigenstate corresponding to the zero eigenvalue shows the notion of the equilibrium state. The gap between the zero eigenvalue and the nearest negative (real) eigenvalue is defined as the asymptotic decay rate (ADR), and the inverse of ADR is denoted as the relaxation time of the system \cite{albert_symmetries_2014}. 

The emergence of the prethermal state in an OQS is less explored than in the isolated systems. In our previous work, we provided a general framework for the emergence of prethermalization in the context of a driven dissipative dipolar coupled systems \cite{saha_cascaded_2023}. For an integrable OQS, the Liouvillian must have multiple zero eigenvalues, which implies that there exist several operators along with the identity matrix (trace preservation), whose expectation values are conserved \cite{albert_symmetries_2014}. In contrast, for a non-integrable OQS, $\mathcal{\hat{L}}$ has a single zero eigenvalue. Similarly, for a nearly-integrable OQS, $\mathcal{\hat{L}}$ consists of the integrable Liouvillian and a perturbation Liouvillian. In such cases, the total $\mathcal{\hat{L}}$ has a single zero eigenvalue. However, there exist several eigenstates whose eigenvalues are very close to zero, which implies that the ADR is very small and the relaxation time is very large. Such a shorter decay rate of the system leads to a prethermal plateau where the dynamics are constrained by the quasi-conserved quantities, which are associated with the eigenstates corresponding to the nearly zero eigenvalues of $ \hat{\mathcal{L}}$. From a practical side, the slowed relaxation of the system has been utilized as a potential quantum storage, as described below.

The use of a spatially-correlated bath for qubit
manipulations has been explored in the last two decades. While the spatio-temporal correlation of a set of interacting particles has been known since the seminal work
of van Hove \cite{van_hove_correlations_1954}. McCutcheon \etal considered two spatially
separated spins coupled to a collection of bosonic modes which acted as the bath
\cite{mccutcheon_long-lived_2009}. Jeske \etal laid out the theoretical framework for a spatially-correlated bath. They considered two
physically separated qubits coupled to a thermal bath \cite{jeske2013}. Two different
formulations of the bath were considered for their analysis. In the first case, a one-dimensional Ising model has been considered as a bath.
For this case, the dephasing rates of the qubits were shown to depend exponentially on the separation of the
two spins. In the second case, a tight-binding chain of the harmonic oscillators has been considered
as the bath. They showed that the spectral densities of the qubits contained a spatial correlation function
that could be modeled as an exponential or a Gaussian form. Complete positivity of their Lindbladian is
ensured for these functional forms. As one expects intuitively, the characteristic correlation length is inversely proportional to the temperature. We note that such spatially-correlated bosonic bath can give
rise to a persistent entanglement to a qubit pair
\cite{braun_creation_2002, benatti_environment_2003, Contreras-Pulido_2008, mccutcheon_long-lived_2009, Zell2009}. 

In this work, we show that depending on the behavior of the spatial correlation function of the spatially correlated bosonic bath, an OQS composed of non-interacting atoms or spins can be identified either as a non-integrable or as an integrable or as a nearly-integrable system. Our results imply that in this scenario, cascaded dynamics consisting of a prethermal plateau can be observed for a nearly integrable system. We find the extensive numbers of quasi-conserved quantities, and we obtain the expression for the quasi-stationary state using a generalized Gibbs ensemble. Moreover, we also study the entropy scaling of the prethermal state and compare it with the thermal state.

We organize the manuscript in the following order: in Sec. \ref{des}, we give a brief description of the system, where we consider a collection of non-interacting atoms coupled with a
spatially correlated bath. We also provide the form of a quantum master equation (QME)
for describing the dynamics of the system. The dynamics of the single spin system are analyzed in Sec. \ref{single-q}. For two spin systems, the dynamical equation is shown in Sec. \ref{two-q}, where all the cases of integrability (i.e., integrable, non-integrable,  and nearly-integrable) are briefly discussed. The emergence of bath-induced entanglement is also discussed in the section. We further generalize our analysis for many-atom systems in Sec. \ref{generalization}. In Sec. \ref{entropy}, we show the behavior of von Neumann entropy. Finally, we briefly discuss the results and their implication in the quantum storage problem in Sec. \ref{discussions} and conclude in Sec. \ref{conclusion}.

\section{Description of the system}
\label{des}
 A general description of a spatially-correlated bosonic bath was provided by Jeske \etal  \cite{jeske2013}. Following in their footsteps,
We consider a collection of $N$ non-interacting qubits coupled to such kind of bath.
 The total Hamiltonian for these systems and the bath is given by
\begin{eqnarray}
\Hq =\sum\limits_{i=1}^N\Hq_{s,i}^\circ +\Hq_{L}^\circ+\hsl.
\label{eq:1}
\end{eqnarray}
Here, $\Hq_{s, i}^\circ$ is the Zeeman Hamiltonian of the individual qubits. $\Hq_{s,i} =
\wo \sigma_z^{i}/2$, where, $\wo$ is the Larmor frequency and $\sigma_i$s are the
Pauli spin matrices. $\Hq_{L}^\circ$ is the free Hamiltonian of the bosonic bath, given by 
$\Hq_{L}^\circ=\sum\limits_k \omega_k a^{\dagger}_ka_k$. $k$ is the number of oscillators in the bath. The qubits are assumed to be weakly coupled to the bath. We
use the coupling Hamiltonian $\hsl$ adapted from the Jaynes–Cummings  model
\cite{Breuer_2002}. As such, we have,
\begin{eqnarray}
\hsl = \sum\limits_{m=1}^N
(\sigma^m_+\mathcal{L}_m +\sigma^m_-\mathcal{L}^{\dagger}_m ), 
\end{eqnarray} 
where $\mathcal{L}_m$ is the
corresponding bath mode. 
The bath modes are defined as, $\mathcal{L}_m(r)= \sum\limits_k g_k^m
a_k$. Here, $g_k^m$ is the coupling amplitude of the $m$th qubit and the bosonic bath and is given by
$g_k^m= g_ke^{ikr_m}$. Hence, $g_k^m \neq g_k^n$ but $\vert g_k^m \vert = \vert g_k^n \vert  $, $\forall m \neq n$.

The dynamical equations of the reduced density matrix of the qubits are given by
the coarse-grained Born-Markov quantum master equation (QME) \cite{Breuer_2002}. In the interaction representation $(\hsl^I)$ \emph{w.r.t} $ \hs^\circ +  \hl^\circ$, such equation is written as,
\begin{eqnarray}
\frac{d \rh(t)}{dt} = -i\tr_L[\hsl^I(t), \rh(t)\otimes \rl]^{\rm sec} -\int\limits_0 ^\infty d\tau \tr_L [\hsl^I(t), [ \hsl^I(t-\tau), \rh(t)\otimes \rl]]^{\rm sec}.
\label{qme-1}
\end{eqnarray}
Here $\rh(t)$ is the system density matrix. $\rl$ is the equilibrium bath density matrix, which is represented by $\rl = e^{-\beta \hl^\circ}/\mathcal{Z_L}$. Here $\beta$ is the inverse temperature of the bath, and $\mathcal{Z_L}$ is the partition function. `$\rm sec$' denotes the secular approximation, which is analogous to the RWA (rotating wave approximation) in quantum optics and nuclear magnetic resonance (NMR). The initial correlation between the system and bath is neglected following the Born-Markov approximation \cite{atom-photon}. The bath is assumed to be isotropic, which means no first-order contribution appears from $\hsl^I$, $\tr_L[ \hsl\rl] = 0$. Such a condition ensures that the system-local environment coupling contributes only in the second order. 

The form of the bath-correlation functions are given by,
\begin{eqnarray}
A_{mn}(\omega_{\circ},r_{mn}) &=& \Re \left[\sum\limits_{k} \vert g_k^2 \vert  \int_0^{\infty}  d\tau \,  e^{-i(\omega_{\circ}-
\omega_k)\tau}\tr_L(a^\dagger_k a_k \rl) e^{-i  k r_{mn} }\right]  \nn\\
B_{mn}(\omega_{\circ},r_{mn}) &=& \Re \left[ \sum\limits_{k} \vert g_k^2 \vert  \int_0^{\infty}  d\tau\,  e^{i(\omega_{\circ}- 
\omega_k)\tau}\tr_L(a_k  a^\dagger_k \rl)
e^{+i  k r_{mn}}  \right]
\end{eqnarray}
Here, $r_{mn} = r_m-r_n$. We only consider the real part of the bath correlation function, as the imaginary part contributes to the Lamb shift, which causes a negligible shift to the Zeeman Hamiltonian \cite{atom-photon}. We also note that, $\tr_L[a^\dagger_m a_m \rl] = N(\omega_m)$,  and $
\tr_L[a_m a^\dagger_m \rl] = 1 + N(\omega_m)$. Here, $N(\omega_m)= \frac{1}{e^{\beta \omega_m}-1}$. We also define $\omega_k = v k_m$. For photon bath, $v = c$. 

  Jeske \etal considered a one-dimensional chain of spatially located, coupled harmonic
oscillators, and the spatial
correlations decay over a characteristic correlation length $\xi = 2\omega_k \beta g_b$. Here, $\omega_k$ is the bath lattice spacing,
$g_b$ is the coupling between the neighboring bath oscillators, and $\beta$ is the inverse temperature. Both the temporal and spatial bath correlations are assumed to be stationary in time and
space. It is expected that when $r_{mn}/\xi \gg 0$, such correlations function $B_{mn}(\omega_{\circ},r_{mn})$, $A_{mn}(\omega_{\circ},r_{mn})$ goes to zero due to the averaging over the highly oscillating phase factor, $\exp(\pm i k r_{mn})$. When the atoms are very close to each other in comparison to the bath correlation length, $r_{mn}/\xi \to 0$, then $B_{mn}(\omega_{\circ},r_{mn}) = B(\omega_\circ)$ and $A_{mn}(\omega_{\circ},r_{mn}) = A(\omega_\circ)$. The detailed calculation for $A(\omega_\circ)$ and $B(\omega_\circ)$ is presented in the appendix \ref{appendix-1}. Their analytical form is given by, $B(\omega_\circ) = \gamma_\circ(1 + N(\omega_\circ))$, and $A(\omega_\circ) = \gamma_\circ N(\omega_\circ)$. Here, $\gamma_\circ = \vert g_k \vert^2 \omega_\circ^2/v^3$. 

Following the previous works by Jeske, McCutcheon \etal, the bath-correlation function can be modeled as, $B_{mn}(\omega_{\circ},r_{mn}) = B(\omega_\circ) \alpha(r_{mn}/\xi)$ and $A_{mn}(\omega_{\circ},r_{mn}) = A(\omega_\circ) \alpha(r_{mn}/\xi)$ \cite{mccutcheon_long-lived_2009,jeske2013}.  Here, $\alpha(r_{mn}/\xi)$ is defined as the spatial correlation function. We note that, $0\leqslant\alpha(r_{mn}/\xi)<1$. For $r_{mn}/\xi \gg 0 $, we get, $\alpha(r_{mn}/\xi) \to 0$, whereas, for $r_{mn}/\xi \to 0$, we get, $\alpha(r_{mn}/\xi) \to 1$. Therefore, $ \alpha(r_{mn}/\xi)$  can be modeled as a decaying function of $r_{mn}$ with a characteristic length scale $\xi$. Depending on the particular model of the bath, such a function can be modeled using a Lorentzian $\xi^2/(\xi^2 + \omega^2 r_{mn}^2)$, or exponential decay function $e^{-r_{mn}/\xi}$, or a Gaussian function $e^{-r_{mn}^2/\xi^2}$ \cite{jeske2013}. For the next part of the analysis, we use $A$, $B$,  and $\alpha$ instead of $A(\omega_\circ)$, $B(\omega_\circ)$, and $\alpha(r_{mn}/ \xi)$ respectively. Using the QME (Eq. (\ref{qme-1})),  the dynamical equation for the $N$ spin system coupled to a spatially correlated Bosonic bath is given as
\begin{eqnarray}
\frac{d \rh}{dt} &=&  \sum\limits_{i=1}^N \left[ B(2\sigma_-^i\rh \sigma_+^i- \{\sigma_+^i\sigma_-^i, \rh\}) + A (2\sigma_+^i\rh \sigma_-^i- \{\sigma_-^i\sigma_+^i, \rh\}) \right]\nn\\
&&+ \sum\limits_{i,j=1}^N \left[  \alpha B(2\sigma_-^i\rh \sigma_+^j- \{\sigma_+^j\sigma_-^i, \rh\}) + \alpha A (2\sigma_+^i\rh \sigma_-^j- \{\sigma_-^j\sigma_+^i, \rh\}) \right], \quad [i\neq j]
\label{n-spin-eq}
\end{eqnarray}
Here, $\sigma_m^i$ represents the Pauli matrix for the $i$th spin. The first term is the self-term ($m=n$), so the spectral densities are only denoted by $A$, $B$. The last term denotes the cross terms between different spins $m \neq n$, so the spectral functions are multiplied by $\alpha$. We note that $\alpha$ is the same for all pairs of atoms as $\vert r_m - r_n \vert$ are assumed to be equal for simplicity.

\section{Dynamics of a single spin system} 
\label{single-q}
If we consider a single spin ensemble in such a spatially correlated bosonic bath, in the dynamical equation (Eq. (\ref{n-spin-eq})) only the first two terms survive, and no cross terms exist. Hence, $\alpha$ has no role in the dynamics. The homogeneous dynamical equation for $N=1$, is given by,
\begin{eqnarray}
\frac{d \rh}{dt} &=&   \left[ B(2\sigma_-^1\rh \sigma_+^1- \{\sigma_+^1\sigma_-^1, \rh\}) + A (2\sigma_+^1\rh \sigma_-^1- \{\sigma_-^1\sigma_+^1, \rh\}) \right]\
\label{1-spin-eq}
\end{eqnarray}
The decomposition of $\rh(t)$ in terms of Pauli spin matrices is given by,
\begin{eqnarray}
\rh(t) = \mathds{I}/2 + \sum\limits_{i=x,y}^{z}\frac{M_i(t)}{2} \sigma_i
\end{eqnarray}
$M_i(t)$ is called the observables. In terms of the observables, the dynamical equation is written as,
\begin{eqnarray}
\dot{M}_x &=& -R_1 M_x \nn\\
\dot{M}_y &=& -R_1 M_y \nn\\
\dot{M}_z &=& -2R_1 M_z  + 2M_\circ R_1
\label{bloch-i}
\end{eqnarray}
 Here we define, $R_1(\omega_\circ) = A(\omega_\circ)+B(\omega_\circ)$ and $M_\circ = \frac{B(\omega_\circ)-A(\omega_\circ)}{B(\omega_\circ)+A(\omega_\circ)}$. In terms of $\beta$, we get $R_1(\omega_\circ) = \gamma_\circ(1 + 2N(\omega_\circ))= \gamma_\circ \coth (\beta \omega_\circ/2)$, and $M_\circ = \tanh(\beta \omega_\circ/2)$. The steady state solution  is given by, $M_x \sts = 0$, $M_y \sts = 0$, and $M_z \sts = M_\circ$. $R_1(\omega_\circ)$ is called the relaxation rate and $M_\circ$ is the equilibrium magnetization. The above equation is an inhomogeneous equation. It was first carried out by Wangsness and Bloch in 1953 \cite{Wangsness1953}. 

Starting from a homogeneous equation, how such inhomogeneity arises due to trace preservation in Eq. (\ref{bloch-i}) is presented in the appendix \ref{appendix-2}. The form of the steady state density matrix is given by,
\begin{eqnarray}
\rh^{\rm eq} &=& \mathds{I}/2 + \frac{1}{2} \tanh(\beta \omega_\circ/2) \sigma_z\nn\\
 \rh^{\rm eq} &=& \frac{e^{-\beta \Hq_{s, 1}^\circ}}{\tr [e^{-\beta \Hq_{s, 1}^\circ}]}
\end{eqnarray} 
 Hence, the steady state solution is a Gibbs state.
 \section{Dynamics of a two spin system}
 \label{two-q}
 The dynamical equation for $N=2$ is given by,
 \begin{eqnarray}
\frac{d \rh}{dt} &=&  \sum\limits_{i=1}^2 \left[ B(2\sigma_-^i\rh \sigma_+^i- \{\sigma_+^i\sigma_-^i, \rh\}) + A (2\sigma_+^i\rh \sigma_-^i- \{\sigma_-^i\sigma_+^i, \rh\}) \right]\nn\\
&&+ \sum\limits_{i,j=1}^2 \left[  \alpha B(2\sigma_-^i\rh \sigma_+^j- \{\sigma_+^j\sigma_-^i, \rh\}) + \alpha A (2\sigma_+^i\rh \sigma_-^j- \{\sigma_-^j\sigma_+^i, \rh\}) \right], \quad [i\neq j]
\label{2-spin-eq}
\end{eqnarray}
The major difference of the dynamical equation for the two spins (Eq. (\ref{1-spin-eq})) from the single spins (Eq. (\ref{2-spin-eq})) is the presence of cross terms, where the spectral densities are multiplied by $\alpha$. The decomposition of $\rh$ for two spins case is given by,
\begin{eqnarray}\label{obs}
\rh &=& \sum\limits_{\alpha, \beta }  A_{\alpha \beta} I_{\alpha} \otimes I_{\beta},
\end{eqnarray}
Here, $\alpha$, $\beta$ can take values from $\{x,y,z,d\}$, and $I_d = 2\times 2$ identity matrix.  Therefore, the density matrix can be written by using sixteen observables. The trace preservation was reduced to fifteen. Such reduction brings the inhomogeneity in the dynamical equation, which is similar to the single spin case. We note that both the atoms are identical. Therefore, the dynamical equation can be further explained by using nine symmetric observables. The contribution from the other six anti-symmetric observables is negligible for the symmetric initial conditions. The representation of the nine symmetric observables is given by,
\begin{eqnarray}
M_{\alpha} &=&   {\rm Tr}_s  [ (I_{\alpha}\otimes \mathbb{I} + \mathbb{I} \otimes I_{\alpha})\rho_s ] \nn\\
M_{\alpha\beta}&=&   {\rm Tr}_s [(I_{\alpha} \otimes I_{\beta} + I_{\beta} \otimes I_{\alpha})
\rho_s ],\quad \forall\ \alpha\neq\beta \nn \\
M_{\alpha\alpha} &=&  {\rm Tr}_s [(I_{\alpha}\otimes I_{\alpha})\rho_s]. 
\label{observables}
\end{eqnarray} 
Here $I_i =\sigma_i/2$. The detailed dynamical equations in terms of observables are presented in the appendix \ref{appendix-3}. We show only those equations of the observables which are non-zero in the steady state. The inhomogeneous dynamical equations are given by,
\begin{equation}\label{3eq}
\left[\begin{array}{c} \dot{M}_z \\ \dot{M}_{zz}\\ \dot{M}_{c} \end{array} \right] = \begin{bmatrix}
-2R_1  & 0 & 4M_{\circ}\alpha R_1 \\   M_{\circ}R_1& -4R_1  & 2 \alpha R_1  \\ 
-M_{\circ} \alpha R_1 & 4 \alpha R_1  & -2R_1  \end{bmatrix}  \left[\begin{array}{c} M_z  \\
M_{zz}\\ M_c \end{array}\right] + \left[\begin{array}{c} 2M_{\circ} R_1 \\  0\\ 0 \end{array}\right]  
\end{equation} 
Here $M_c = M_{xx} + M_{yy}$. For the numerical simulation, the natural choice for analyzing the master equation described by Eq. (\ref{2-spin-eq}), is to adopt a Liouville
space description.  Liouville space form of this QME is, $ \frac{d \hat{\rho_s}}{dt} =
\mathcal{\hat{L}}\hat{\rho_s} $ where, $\mathcal{\hat{L}}$ is the Liouvillian superoperator.  The
resulting Liouvillian matrix is a $N^2 \times N^2$ matrix, and the density matrix is a $N^2 \times 1$ column
matrix, where $N$ is the dimension of the Hilbert space. We note that, despite the simplicity of appearance, the size of the Liouvillian is not convenient
for algebraic manipulation. 

The steady-state solution of QME can be obtained by solving $\mathcal{\hat{L}}\mathcal{\hat{\rho}}_s=0$. The steady-state density matrix is the eigenvector corresponding to the zero eigenvalues of the Liouvillian superoperator. If $\mathcal{\hat{L}}$ has a single zero eigenvalue, then the steady state has no initial value dependence. For such systems, the asymptotic decay rate (ADR) is defined as the gap between the zero eigenvalue and the nearest negative eigenvalue \cite{buca_note_2012}. Its inverse provides the relaxation time of the system. In general, eigenvalues of the Liouvillian are functions of the system parameters $\{q\}$. If, for a particular value of a parameter, the Liouvillian has multiple zero eigenvalues, then it is evident that there exists conserved quantity in the system, which leads to the initial value dependence of the steady state. 

We note that $\alpha$ plays a major role in the steady state behavior. Next, we present the steady state analysis for different values of $\alpha$.
 
 \subsection{Steady state solution for $\alpha<1$}
 For $\alpha<1$, the Liouvillian $\hat{\mathcal{L}}$ has a single zero-eigenvalue. In the above Eq. (\ref{3eq}), the square matrix is also non-singular. The steady-state solution is given by,
 \begin{eqnarray}\label{3sol-neq1}
M_z\sts &=& M_{\circ} \nonumber \\
M_{zz}\sts &=& M_{\circ}^2/4 
\end{eqnarray}
The corresponding density matrix is given by,
\begin{eqnarray}
 \rh = \frac{\mathds{1}}{4} + \frac{M_z}{2} (I_{z}\otimes \mathbb{I} + \mathbb{I} \otimes I_{z})   + 4 M_{zz} I_{z}\otimes I_z
 \label{ge1}
 \end{eqnarray}
 The numerical factors arise in each term due to the division of $\tr(\mathcal{O}^2)$. Here $\mathcal{O}$ is the particular operator.
From the previous calculation, we know that $M_\circ = \tanh \frac{\beta \omega_\circ}{2}$.
Using it, we get
\begin{eqnarray}
\rh^{ss} &=& \frac{\exp{\left( -\sum\limits_{i=1}^2\beta \mathcal{H}_{s,i}^\circ \right)}}{\tr\left[\exp{\left( -\sum\limits_{i=1}^2\beta \mathcal{H}_{s,i}^\circ \right)} \right]}.
\label{ge2}
\end{eqnarray}
Hence, for $\alpha<1$, the steady state is a Gibbs state. The purity $(\tr[\rh^2])$ of the corresponding state is given by,
\begin{eqnarray}
\tr[\rh^2]\vert_{\alpha<1 } &=& \frac{1}{2} \frac{1 + \cosh(2 \beta \omega_\circ)}{(1 + \cosh(\beta \omega_\circ))^2}
\label{p<1}
\end{eqnarray}
In the Fig. \ref{fig-time-temp}, $M_\circ =0.8$, so $\beta \omega_\circ = \ln(9)$. By putting the values in the Eq. (\ref{3sol-neq1}), for the initial condition, $\rh\sto = \mathds{I}/4$, we get $M_z\sts = 0.8$, $M_{zz}\sts = 0.16$, and $M_c =0$, which matches with the numerical results shown in Fig. \ref{fig-time-temp} (a), (b) and (c) for $\alpha <1$.
\subsection{Steady state solution for $\alpha=1$}
For $\alpha=1$, the Liouvillian $\hat{L}$ has  two zero-eigenvalues. We also note that in the above Eq. (\ref{3eq}), the square matrix is singular, which implies that there exists a conserved quantity in the system. The expression of the conserved quantity is given by,
\begin{eqnarray}
\frac{d}{dt}(M_c + M_{zz})=  \frac{d}{dt} \langle \vec{\sigma}_1. \vec{\sigma}_2 \rangle =  0
\label{conserved-1}
\end{eqnarray}
The singularity also implies that the final steady state has an initial value dependence. The steady-state solution in terms of observables is given by,
\begin{eqnarray}\label{3sol-eq1}
M_z\sts &=& M_{\circ} \left[   \frac{ (4 F + 3)}{      
\left(M_{\circ}^2+3\right)} \right] \nonumber \\
M_c\sts &=& \left[\frac{4 F -  M_{\circ}^2  }{2     
\left(M_{\circ}^2+3\right)}\right] \nonumber \\
M_{zz}\sts &=& F - M_c 
\end{eqnarray}
where, $ (M_{zz}+M_{c})\ic = F$. This result is in complete agreement with earlier work \cite{benatti2006b}. The steady-state configuration using the above values is written as,
\begin{eqnarray}
 \rh = \frac{\mathds{I}}{4} + \frac{M_z }{2} (I_{z}\otimes \mathbb{I} + \mathbb{I} \otimes I_{z})   + 4 M_{zz}  I_{z}\otimes I_z + 2M_c (I_x\otimes I_x + I_y \otimes I_{y})  
 \label{ge11}
 \end{eqnarray}
Similarly, the form $\rh^{ss}$ can be written using a generalized Gibbs state, which is given by,
\begin{eqnarray}
\rh^{ss}\vert_{\alpha=1} &=& \frac{\exp{\left(-\sum\limits_{i=1}^2\beta \mathcal{H}_{s,i}^\circ - \sum\limits_{i,j=1}^2 \frac{l_1}{4}\vec{\sigma_i} . \vec{\sigma_j}\right)}}{\mathcal{Z}_g}, \quad[\forall i \neq j].
\label{GGE}
\end{eqnarray}
Where, $\mathcal{Z}_g = \tr\left[\exp{ \left(-\sum\limits_{i=1}^2\beta \mathcal{H}_{s,i}^\circ- \sum\limits_{i,j=1}^2\frac{l_1}{4}\vec{\sigma_i} . \vec{\sigma_j}\right)}\right]$. $\mathcal{Z}_g$ is the partition function. $l_1$ is the Lagrange undetermined multiplier. The value of $l_1$ can be determined from the value of $(M_{zz}+M_{c})\ic = F$, which is written as,
\begin{eqnarray}
l_1 = \ln\left(\frac{1-4F}{3+4F}\times \left(  1 +  2\cosh\beta\omega_{\circ}\right) \right).
\end{eqnarray} 
The purity of the corresponding state is given by,
\begin{eqnarray}
\tr[\rh^2]\vert_{\alpha=1 } &=&  \frac{-2-8F+(5+8F(1+2F))\cosh\beta \omega_\circ}{4+8\cosh \beta \omega_\circ}
\label{p=1}
\end{eqnarray}
For the same initial condition, $\rh\sto = \mathds{I}/4$, we get $F=0$, and we choose $M_\circ =0.8$ by putting the values in Eq. (\ref{3sol-eq1}), the steady state solution is given by, $M_z\sts =  0.6593$, $M_c \sts = -0.0879$, and $M_{zz} = 0.0879$, which matches with the results of the numerical simulations for the $\alpha = 1$ case shown in Fig. \ref{fig-time-temp} (a), (b), and (c).
\subsection{Steady state solution for $\alpha \to 1$}
The condition $\alpha \to 1$ is physically more achievable than $\alpha =1$ in experiments. For $\alpha \to 1$, although the Liouvillian $\mathcal{\hat{L}}$ has a single zero eigenvalue, the corresponding ADR $ \to 0$, this implies that the relaxation time increases sufficiently. As such, the system reaches a quasi-steady state in the intermediate time scale, and the corresponding quasi-steady state and the dynamics are constrained by a quasi-conserved quantity in this time scale. After a very long time, the system reaches the thermal state. We also provide the results of the numerical simulations to justify the existence of the quasi-steady state.
\begin{figure}[htb]
\raisebox{3cm}{\normalsize{\textbf{(a)}}}\hspace*{-1mm}
\includegraphics[width=0.42\linewidth]{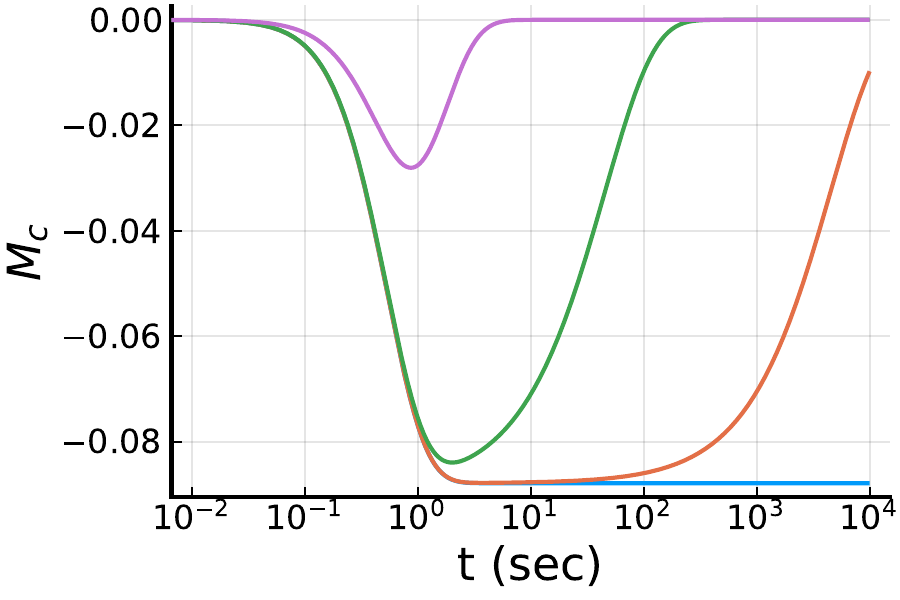} 
\hspace*{4mm}
\raisebox{3cm}{\normalsize{\textbf{(b)}}}\hspace*{-1mm}
\includegraphics[width=0.42\linewidth]{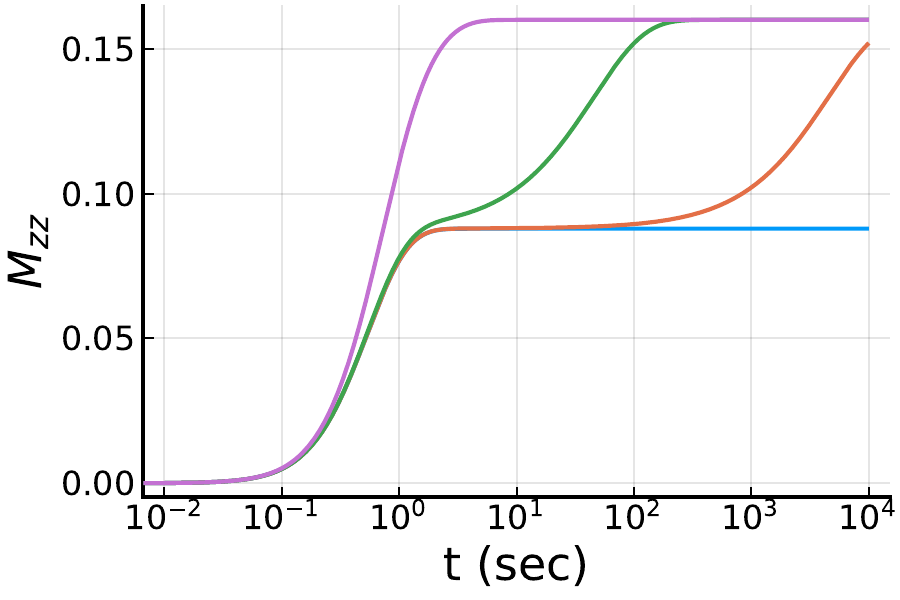}
\hspace*{4mm}
\raisebox{3cm}{\normalsize{\textbf{(c)}}}\hspace*{-1mm}
\includegraphics[width=0.42\linewidth]{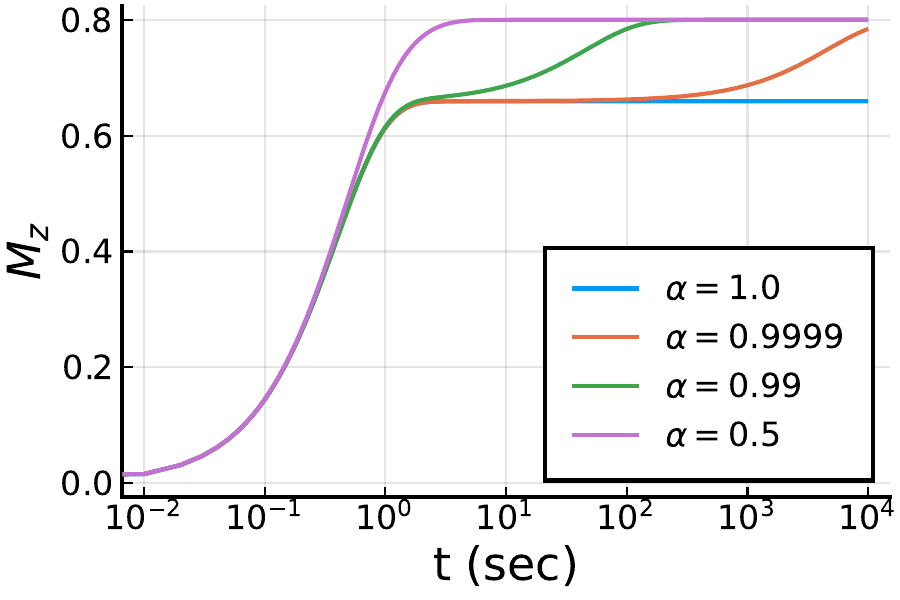} 
\hspace*{4mm}
\raisebox{3cm}{\normalsize{\textbf{(d)}}}\hspace*{-1mm}
\includegraphics[width=0.42\linewidth]{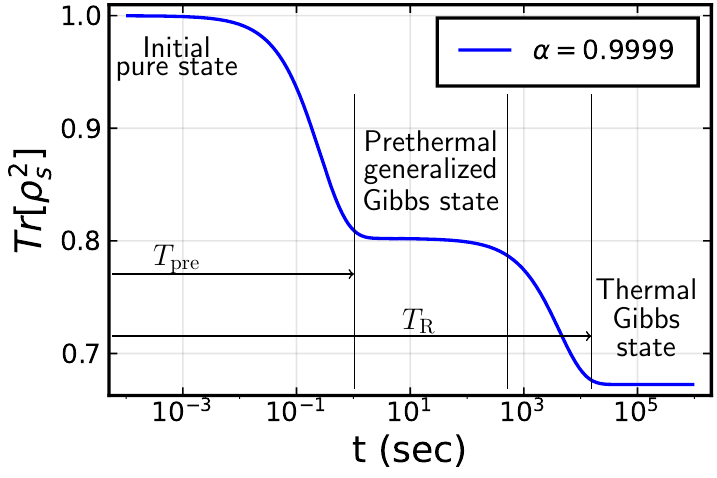} 
\caption{Figures (a), (b) and (c) shows the plots of three observables $\{M_c, M_{zz}, M_c\}$ as a function of time
$(t)$ respectively, obtained by numerically solving Eq. (\ref{3eq}), for a set of values of $\alpha$,
[$\alpha = 1.0$ (blue, color online; the lower curve), $\alpha = 0.9999$ (orange, color online;  the second from the bottom), $\alpha = 0.99$ (green, color online; the second from the top), $\alpha = 0.5$
(pink, color online; the upper curve)]. The fixed parameter sets are chosen as $\{R_1 = 1.0, M_{\circ} = 0.8 \}$. The initial value is $\{ M_z \sto = M_{zz}\sto = M_c\sto =0\}$. For $\alpha =
1.0$, the steady state values of all three observables are significantly different from $\alpha < 1$ and
are given by Eq. (\ref{3sol-eq1}). For $\alpha \neq 1$, the observables reach the thermal equilibrium
values. For $\alpha$  close to 1, a slowed relaxation of the three observables is noticed as there exists a prethermal plateau. Fig. (d) shows the plot of $ \tr \left[ \rh^2 \right]$ as a function of time for $\alpha = 0.9999$. The initial state is chosen as $\vert\psi_{ini} \rangle=\vert \uparrow \uparrow\rangle$. At a time $T_{\rm pre}$, the prethermal phase occurs, and it remains as a plateau until $T_{\rm R}$, after which it further decays at a time $T_{ 1}$ to the thermal state.   } 
\label{fig-time-temp}
\end{figure}
 Prethermalization is defined as the partial loss of the initial memory of the system in the intermediate timescale \cite{ueda_2020,marino_relaxation_2012,kollar_2011,abanin_rigorous_2017}. It is a quasi-steady state which appears in the intermediate timescale. An important characteristic of the prethermal state is the presence of at least one quasi-conserved quantity in the system. By quasi-conserved quantity, we indicate a quantity that appears to be conserved for the intermediate prethermal state but not in the final equilibrium. 
The time-series plots of the observables $\{M_z, M_c, M_{zz}\}$for $\alpha =0.9999 $ (Fig. \ref{fig-time-temp}) show the notion of multiple timescales in the system. The observable in Eq. (\ref{conserved-1}) acts as a quasi-conserved quantity in the system. We notice the system follows the dynamics for $\alpha=1$ in the intermediate timescale, whereas it follows the dynamics $\alpha<1$ at the final time. Hence, the system follows the generalized Gibbs distribution in the prethermal phase, and it inherits the Gibbs distribution at the final equilibrium. In Fig. \ref{fig-time-temp} (a), (b), and (c), for $\alpha = 0.9999$, all the observables reach the prethermal state, whose expectation values match with $\alpha=1$, at larger time, such values are equal to $\alpha<1$ case.

 We also provide a numerical plot of purity ($\tr[\rh^2]$) versus time for $\alpha = 0.9999$ in Fig. \ref{fig-time-temp}(d). The plot shows cascaded dynamics. For an initially prepared pure state $(\vert \psi_{ini} \rangle=\vert \uparrow\uparrow \rangle)$, the system reaches the prethermal state at a time $T_{\rm pre}$, and the purity decreases to a quasi-equilibrium value. At an approximate time $T_{\rm R}$, the purity further decreases as the system reaches the thermal state. The lifetime of the prethermal plateau increases by increasing the value of $\alpha$. At $\alpha=1$, the prethermal phase becomes a stable phase. For such a choice of initial state, we get $F = \frac{1}{4}$. Here, $M_\circ =0.8$, so $\beta \omega_\circ = \ln(9)$. Putting the values of $F,\, \beta\omega_\circ$ in Eq. (\ref{p=1}) and Eq. (\ref{p<1}), we get $\tr[\rh^2] = 0.8022$ at $\alpha =1$ and  $\tr[\rh^2] = 0.6724$ at $\alpha <1$. The numerical result of purity (shown in Fig. \ref{fig-time-temp}(d)) matches with the analytical result.
\subsection{Measure of Entanglement}
We note that $M_c = \rho_{32}+\rho_{23}$, which corresponds to the zero-quantum coherence
of the system. The presence of zero-quantum coherence at the steady state for $\alpha=1$ signifies that there is a possibility of having quantum entanglement between the spin pairs. In the case of a bipartite system, concurrence is a measure of entanglement.
It was defined using the eigenvalues of the following Hermitian matrix $\sqrt{\sqrt{\rh } \tilde{\rh },
\sqrt{\rh }}$. Here, $\rh $ is the system density matrix. Also, $\tilde{\rh }=
\sigma_y\otimes\sigma_y\, \rh ^{\star}\,\sigma_y \otimes \sigma_y$. Where $\star$ indicates the complex conjugate. Similarly, it can be
calculated from the square root of eigenvalues of the non-Hermitian matrix $\rh  \tilde{\rh } $. The concurrence $(C(\rh  ))$ is defined as,
$C( \rh  ) = \text{max}\{0, \lambda_1-\lambda_2-\lambda_3-\lambda_4\}$ \cite{hill_entanglement_1997}.
Here, $\{\lambda_i\}$ are the eigenvalues arranged in decreasing order. For $C( \rh  )= 0$, the system is separable, whereas if $C( \rh  )> 0$, there exists entanglement in the system. In terms of $M_z$, $M_{zz}$, and $M_c$, the eigenvalues $\lambda_i$ are given by,
\begin{eqnarray}
\lambda_i &=& \{ \frac{1}{4} \sqrt{(1 + 4 M_c -4 M_{zz})^2},\, \frac{1}{4} \sqrt{(-1 + 4 M_c +4 M_{zz})^2},\,\nn\\
&& \frac{1}{4} \sqrt{(1 + 4 M_{zz})^2 - 4M_z^2},\, \frac{1}{4} \sqrt{(1 + 4 M_{zz})^2 - 4M_z^2} \}.
\end{eqnarray}
Arranging them in descending order the expression of $C( \rh  )$ is given by,
\begin{eqnarray}
C( \rh  ) &=& \frac{1}{2}\left( 4 \vert M_c \vert - \sqrt{(1 + 4 M_{zz})^2 - 4M_z^2} \right).
\end{eqnarray} 
For $\alpha <1$, $M_c = 0$, the final steady state is a separable one. On the other hand, for $4 \vert M_c \vert>\sqrt{(1+4M_{zz})^2-4M_z^2}$,  there is a possibility of having the bath-induced persistent entanglement at $\alpha = 1$. We note here that for $\alpha \to 1$, the existence of long-lived entanglement was previously reported by McCutcheon \etal\cite{mccutcheon_long-lived_2009}.

\section{Dynamics of N-spin system}
\label{generalization}
In this section, we provide the generalization of the two-spin problem discussed above. We find that, for $\alpha<1$, the Liouvillian $\mathcal{\hat{L}}$ has a single zero eigenvalue for any values of $N$. The number of observables for finding the steady state will increase with $N$. For example, in the appendix \ref{appendix-4}, we present the dynamical equation of the relevant observables for three spin systems, where the number of observables is five $\{M_z,\, M_{zz},\, M_{zzz}, \, M_{xx} + M_{yy},\, M_{xxz} + M_{yyz}\}$. The analytical complications can be reduced by looking at the properties of the steady state. We note that for $\alpha =0$ and $0<\alpha<1$, the steady-state behavior is the same for any values of $N$. We also verify the above statement numerically by solving Eq. (\ref{n-spin-eq}) upto five spins for $0\leqslant\alpha<1$. Therefore, in case of $\alpha<1$, such steady-state configuration can be thought of as the steady state of a collection of non-interacting spins coupled to an individual local environment, which is given by,
\begin{eqnarray}
\rh^{ss} =  \frac{e^{-\sum\limits_{i=1}^N\beta \mathcal{H}_{s,i}^\circ}}{\mathcal{Z}}
\end{eqnarray}
The above state is Gibb's state. There exist no conserved quantities at $\alpha<1$. On the other hand, the dynamics of the $N$ spin system are markedly different for $\alpha=1$. The number of zero eigenvalues of $\mathcal{\hat{L}}$ increases with increasing the atom number $N$. We numerically find those numbers up to five spins, which are given by 2, 5, 14, and 42 for the no. of atoms are 2, 3, 4, and 5, respectively. The presence of multiple zero eigenvalues ensures that there exist several conserved quantities in the system for higher values of $N$. Here, we are only interested in the conserved quantities corresponding to the two-spin correlation. Using the above Eq. (\ref{n-spin-eq}), we find the expression for the existing conserved quantities.
\begin{eqnarray}
 \frac{d}{dt}\langle \vec{\sigma}_i.\vec{\sigma}_j \rangle = 0, \quad [i<j].
\label{conserved-n}
\end{eqnarray}
The above Eq. (\ref{conserved-n}) is the generalization of Eq. (\ref{conserved-1}). Therefore, for the $N$ spin system, the number of conserved quantity is ${N \choose 2}$ at $\alpha =1$. Such an extensive number of conserved quantities shows the notion of integrability in the system. There exist a global symmetry super-operator $(\mathcal{\hat{O}})$, which commutes with the Liouvillian $[\mathcal{\hat{O}}, \mathcal{\hat{L}}]=0$. The form of the symmetry operator is given by,
\begin{eqnarray}
\mathcal{O} &=& \sum\limits_{i,j =1}^N\frac{\vec{\sigma_i}}{2} . \frac{\vec{\sigma_j}}{2}\nn\\
            &=& \frac{3N}{4} \mathbb{I} + \sum\limits_{i,j =1}^N \frac{\vec{\sigma_i}}{2} . \frac{\vec{\sigma_j}}{2}, \quad[\forall i\neq j]. 
\label{symmetry1}
\end{eqnarray}
We note that the Liouvillian is invariant under the following symmetry transformation. $\hat{U} = \exp{(-i \phi \hat{\mathcal{O}})}$. Here $\phi$ is any real parameter. So, $\hat{U} \mathcal{\hat{L}} \hat{U}^{\dagger} = \mathcal{\hat{L}}$. Following the definition of the symmetry in QME by  Buca, Lieu \etal, a strong symmetry condition is satisfied in this case, as $\mathcal{\hat{O}}$ commutes with the superoperator corresponding all Hamiltonians and Lindbladian jump operators in the QME. On the other hand, a weak symmetry is satisfied when it only commutes with $\mathcal{\hat{L}}$ but not with its individual components \cite{lieu_symmetry_2020,buca_note_2012,albert_symmetries_2014}.

The consequence of having a strong symmetry in the system is the following. As this symmetry operator is the total angular momentum operator, therefore, in the collective basis or angular momentum basis, $\hat{\mathcal{L}}$ can be written in the block diagonal form at $\alpha =1$. Each block will contain at least one of the zero eigenvalues, which gives the notion of the reduced subspace of the final steady state \cite{minganti_spectral_2018}. In this case, the corresponding global conserved quantity is the sum of all local conserved quantities. Those eigenstates corresponding to the operator $\mathcal{O}$, which are not evolving under $\mathcal{\hat{L}}$ are known as the dark states or decoherence-free subspaces (DFS) \cite{buca_note_2012}.

 The collective operator is defined as $J_i = \sum_i \sigma_i$, and the corresponding eigenbasis is given by $\vert J M \rangle$. Here, $J_{max}=\frac{N}{2}$ and $M$ can be varied from $-N$ to $N$. The value of $J$ can be varied from $1$ to $\frac{N}{2}$. Therefore, the number of independent blocks for $N$ spin system is $\frac{N}{2}+1$, if $N$ is even, otherwise it is equal to, $\frac{N+1}{2}$. For $\alpha =1$, the Eq. (\ref{n-spin-eq}) in terms of collective operators is written as,
 \begin{eqnarray}
 \frac{d \rh}{dt} &=&    B(2J_-\rh J_+ - \{J_+  J_- , \rh\}) + A (2J_+ \rh J_- - \{J_-J_+, \rh\})
\label{parental1}
\end{eqnarray}
 In a collective basis, the observables for the particular $J$ block are given by,
  \begin{eqnarray}
P_{M M^\prime}^J = \tr[\rh \vert J M \rangle \langle J M^\prime \vert]
\end{eqnarray}
Using Eq. (\ref{parental1}), we find that there exists a constraint, $\sum_M P_{M M }^J = 1$. Such constraint leads to inhomogeneity in the system, which implies that only diagonal elements survive and the off-diagonal elements vanish at the steady state. We will use $P_M$ in place of $P_{MM}^J$ for the next part of the analysis. For $J=\frac{N}{2}$ block, the number of independent observables is $N$. The QME (Eq. \ref{parental1}) in terms of observables for the particular block is given by,
\begin{eqnarray}
\frac{d P_M}{dt} &=& -\zeta(M)(A(0)P_M - B(0) P_{M+1})- \zeta(-M)(B(0) P_M - A(0) P_{M-1})
\label{nspin-sol}
\end{eqnarray}
Here, $\zeta(M) = 2(J-M)(J+M+1)$ and $M$ can be varied from $-J$ to $J$. For the two spins case, the solution of the above Eq. (\ref{nspin-sol}) is in agreement with Eq. (\ref{3eq}). More details on the exactness of the two Eq. (\ref{nspin-sol}) and (\ref{3eq}) at $\alpha=1$ can be found in the appendix \ref{appendix-5}.

According to Eq. (\ref{nspin-sol}), every block reaches a steady state when the following condition is satisfied,
\begin{eqnarray}
\frac{P_M}{P_{M-1}}  &=& \frac{A(0)}{B(0)} = \exp{\left(-\beta \omega_{\circ}\right)}.
\label{n-conditon}
\end{eqnarray}
Hence, the $J^{\rm th}$ block of the steady-state density matrix can be represented as,
\begin{eqnarray}
\rho_{J}^{ss}\sts = \frac{1}{\tr[\rho_J\sto]} \times\frac{\exp{ \left(-\beta \omega_{\circ}J_Z \right)}}{\tr\left[\exp{\left(-\beta \omega_{\circ}J_Z \right)}\right]}
\end{eqnarray}
where, $\rho_{J}$ is the representation of the $J^{\rm th}$ block of $\rh$ and $J_Z = M \vert J M\rangle \langle J M \vert$. 
 \setlength{\arraycolsep}{0pt}
The explicit block-diagonal form of $\rh^{ss}$ is given by,
  \begin{eqnarray}
\rh^{ss}(\alpha=1) &=& \begin{bmatrix}
 \boxed{\rho_{J}^{ss} } &   &   &    \\   &  \boxed{\rho_{J-1}^{ss}} &    &      \\     &   & \ddots &  &  \\ & & &  \boxed{\rho_{0}^{ss}  } \end{bmatrix}
\label{new_steady}
\end{eqnarray}
 
The above solution in Eq. (\ref{new_steady}) agrees well with the earlier work of Latune \etal \cite{Latune_2019}.
The solution shows that every sub-space of the density matrix gets the Gibbs distribution. We note that such separate thermalization of each block is a consequence of GGE \cite{halati_2022}. Along with the complete trace preservation, we find that the trace of each sub-space also remains preserved at $\alpha =1$. Hence, at the steady state, the total initial population of each $J$ block remains unchanged. 
\section{Measure of Entropy} 
\label{entropy}
To further analyze the inherent integrability of the system, we calculate the von Neumann entropy. The expression of the von Neumann entropy, for a quantum mechanical system, is defined as, 

\begin{eqnarray}
\mathcal{S} =  - \tr[\rh \ln \rh]. 
\end{eqnarray}
Here, $k_B =1$. The analytical form of $\mathcal{S}$ can be directly calculated from the partition function $\mathcal{Z}$. The average energy $U$, in terms of $\beta$ and $\mathcal{Z}$ is given as, $U = -\frac{1}{\mathcal{Z}} \frac{\partial \mathcal{Z}}{\partial \beta}$. From the thermodynamics relations, we know that $\mathcal{S}=\beta U + \ln \mathcal{Z}$. Hence, in terms of $\beta$ and $\mathcal{Z}$, the entropy $(\mathcal{S})$ can be expressed as,
\begin{eqnarray}
\mathcal{S} = -\beta^2 \frac{\partial}{\partial \beta} \left(\frac{\ln \mathcal{Z}}{\beta}\right)
\end{eqnarray}
 We study the behavior of $\mathcal{S}$ as a function of $N$ for two choices of $\alpha$ ($\alpha <1$, and $\alpha =1$).
\subsection{Entropy for $\alpha<1$}
\begin{figure*}[htb]
\includegraphics[width=0.4\linewidth]{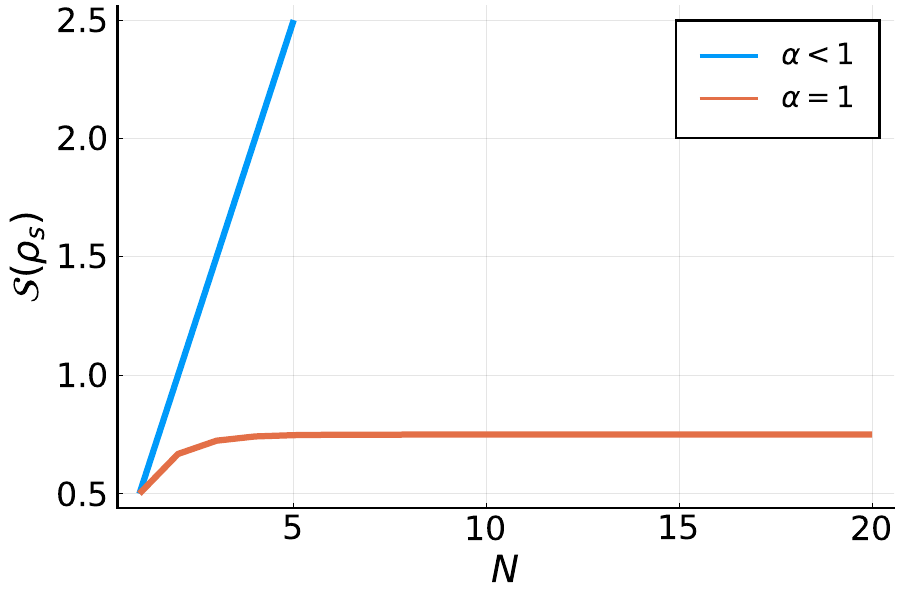}
\caption{The figure shows the plot of steady state von Neumann entropy in the units of $k_B$ as a function of the number of spins. The initial state is chosen as, $\vert \psi \rangle \sto = \vert \downarrow \downarrow ... \downarrow_N \rangle$, or $[\vert \psi \rangle \sto = \vert  \frac{N}{2}, \frac{N}{2} \rangle]$, so the dynamics is confined in the principal $J$ block. The values of the parameters are given below, $R_1 = 1.0$, $M_{\circ}=0.6$. For $\alpha < 1 $ (sky-blue color online; the upper plot), entropy is an extensive quantity and it follows the volume law (i.e., increases linearly with increasing the number of atoms). On the other hand, for $\alpha = 1 $ (orange color online; the lower plot ), the entropy becomes constant for large $N$. It is not an extensive quantity anymore.  }
\label{fig-4}
\end{figure*}

The analytical form of $\mathcal{Z}$, for $\alpha<1$ is given as, $\mathcal{Z}= \tr[e^{-\sum\limits_{i=1}^N\beta \mathcal{H}_{s,i}^\circ}]$. After rearranging, such term can be written as, $\mathcal{Z}= \tr[e^{-\beta \mathcal{H}_{s,1}} \otimes e^{-\beta \mathcal{H}_{s,2}}  \otimes .... \otimes e^{-\beta \mathcal{H}_{s,N}}]$. Using the property of the trace operator, $\tr(A\otimes B)= \tr(A)\times \tr(B)$, the final form of $\mathcal{Z}$ is given by,
\begin{eqnarray}
\mathcal{Z}= \left(\tr[e^{-\beta \mathcal{H}_{s,1}}]\right)^N
\end{eqnarray}
For given $\beta,\, \omega_\circ$, the entropy is written as,
\begin{eqnarray}
\mathcal{S}_{\alpha<1} &=&  N\left(\ln \left(2 \cosh\left(\frac{\beta \omega_\circ}{2} \right) \right) - \frac{\beta \omega_\circ}{2} \tanh\left(\frac{\beta \omega_\circ}{2} \right) \right)
\label{entropy<1}
 \end{eqnarray}
Hence, for $\alpha <1$, the entropy explicitly depends on $N$. For $M_\circ =0.6$, we get $\beta \omega_\circ = \ln(4)$. Putting the value in Eq. (\ref{entropy<1}), we get $\mathcal{S}_{\alpha<1} = 0.5004 N$, which matches with the numerical solution (shown in Fig. \ref{fig-4} (sky-blue color online)), we get by solving Eq. (\ref{n-spin-eq}) for $\alpha <1$ up to five spins.
\subsection{Entropy for $\alpha=1$}
It is evident that, for $\alpha =1$, due to the presence of extensive numbers of conserved quantities, the behavior of $\mathcal{S}$ would be different from $\alpha =1$. The dynamics is confined in a particular $J$ block, so we provide a systematic study of $\mathcal{S}$, for the different choice of the initial population in the particular $J$  blocks, which may be non-degenerate, degenerate, or may be the combinations of both.
\subsubsection{Dynamics confined in the non-degenerate $J$ block}
For this case, we assume that the initial population is distributed in the principal block $J=\frac{N}{2}$. We note that the principal block $J=\frac{N}{2}$ is non-degenerate. The partition function for this case is given as, $\mathcal{Z}_N =   \tr[\exp(-\beta \omega_\circ J_z)]$, which is expressed as,  
\begin{eqnarray}
\mathcal{Z}_N = \frac{ \sinh \left( \frac{N+1}{2} \beta \omega_\circ \right)}{\sinh \frac{\beta \omega_\circ}{2} }.
\end{eqnarray}
We note that, for  $N\gg1$, $\ln \mathcal{Z}_N = N \frac{\beta \omega_\circ}{2} - \ln(2\sinh \frac{\beta \omega_\circ}{2})$. Using the formula for $\ln \mathcal{Z}_N$, the expression of $\mathcal{S}$ is given by,
 \begin{eqnarray}
\mathcal{S}_{\alpha=1} &=& \frac{\beta \omega_\circ}{2}  \coth\left( \frac{\beta \omega_\circ}{2 } \right)- \ln\left(2 \sinh\left( \frac{\beta \omega_\circ}{2} \right) \right).
\label{entropy=1}
 \end{eqnarray}
 Therefore, for $N \gg 1$, the entropy saturates and it doesn't depend on $N$. For the initial state, $\vert \psi \rangle \sto = \vert \frac{N}{2}, -\frac{N}{2} \rangle$, if we choose $\beta \omega_\circ = \ln(4)$, the value of $\mathcal{S}$ from Eq. (\ref{entropy=1}) is given by, $\mathcal{S} = 0.74798$, which matches with the numerical results, we get from Eq. (\ref{nspin-sol}). The result of the numerical simulation is shown in Fig. \ref{fig-4}  (orange color online).
\subsubsection{Dynamics confined in the degenerate $J$ block}
We note that, for $N>2$, all the other blocks excluding the principal block  $(J<\frac{N}{2})$, are degenerate, such degeneracy factor is given by \cite{mandel995},
\begin{eqnarray}
D_l = \frac{( 2l +1) N!}{(\frac{N}{2}+l +1)! (\frac{N}{2}-l)!}, \quad[l \leqslant \frac{N}{2}].
\end{eqnarray}
The $J^2$ symmetry implies that, for the initial population in any one of the degenerate blocks, the dynamics are similarly confined in that particular block. However, there is a possibility of mixing between the different degenerate blocks, which could increase the entropy of the system. Such mixing is prohibited if there exist several local conserved quantities in the system. For example, for $N=3$, $J = \frac{3}{2}$, block is non-degenerate and $J = \frac{1}{2}$, block is degenerate. Using the above formula, we find that the degeneracy is two, so $\vert \frac{1}{2}, \frac{1}{2} \rangle $ and $\vert \frac{1}{2}, -\frac{1}{2} \rangle $ is double degenerate. The expression of the degenerate states is given by \cite{mandel995},
\begin{eqnarray}
 \vert \frac{1}{2}, -\frac{1}{2}, a \rangle &=& \frac{1}{\sqrt{2}}(\vert \uparrow \downarrow \downarrow \rangle - \vert \downarrow \uparrow \downarrow \rangle),\nn\\
  \vert \frac{1}{2}, -\frac{1}{2}, b \rangle &=& \frac{1}{\sqrt{6}}(\vert \uparrow \downarrow \downarrow \rangle + \vert \downarrow \uparrow \downarrow \rangle - 2\vert \downarrow \downarrow \uparrow  \rangle ),\nn\\
   \vert \frac{1}{2}, \frac{1}{2}, a \rangle &=& \frac{1}{\sqrt{2}}(\vert \downarrow \uparrow \uparrow \rangle - \vert \uparrow \downarrow \uparrow \rangle) \nn\\
  \vert \frac{1}{2},  \frac{1}{2}, b \rangle &=& \frac{1}{\sqrt{6}}(\vert \downarrow \uparrow \uparrow \rangle + \vert \uparrow \downarrow \uparrow \rangle - 2\vert \uparrow \uparrow \downarrow  \rangle ).\nn
\end{eqnarray}
If we choose the initial state as $\vert \frac{1}{2}, \frac{1}{2}, a \rangle $, then in the steady state the  eigenvalues of $\rh \sts$ are $\frac{e^{-\beta \omega_\circ/2}}{e^{-\beta \omega_\circ/2} +e^{+\beta \omega_\circ/2}}$, $\frac{e^{\beta \omega_\circ/2}}{e^{-\beta \omega_\circ/2} +e^{+\beta \omega_\circ/2}}$, and the remaining eigenvalues are zero. Similarly for initial state $\vert \frac{1}{2}, \frac{1}{2}, b \rangle $  the eigenvalues are also same. For their linear combination,   $\alpha_1 \vert \frac{1}{2}, \frac{1}{2}, a \rangle  + \alpha_2 \vert \frac{1}{2}, \frac{1}{2}, b \rangle$  [here, $ (\alpha_1^2 + \alpha_2^2 = 1 )$], the steady state eigenvalues remain intact. The above results imply that the degeneracy of the same $J$ remains intact at $\alpha = 1$. We note that, for $N=3$, there exist three local conserved quantities, viz., $\langle \vec{\sigma}_1. \vec{\sigma}_2 \rangle$, $\langle \vec{\sigma}_2. \vec{\sigma}_3 \rangle$, and $\langle \vec{\sigma}_2. \vec{\sigma}_3 \rangle$. $\vert \frac{1}{2}, -\frac{1}{2}, a \rangle = \frac{1}{\sqrt{2}}(\vert \uparrow_1 \downarrow_2  \rangle - \vert \downarrow_1 \uparrow_2 \rangle) \otimes \vert\downarrow_3\rangle  $, so it is an eigenstate $ \vec{\sigma}_1. \vec{\sigma}_2 $, whereas, $\vert \frac{1}{2}, -\frac{1}{2}, b \rangle = \frac{1}{\sqrt{6}} ((\vert \uparrow_1 \downarrow_3  \rangle - \vert \downarrow_1 \uparrow_3 \rangle) \otimes \vert\downarrow_2\rangle + (\vert \uparrow_2 \downarrow_3  \rangle - \vert \downarrow_2 \uparrow_3 \rangle) \otimes \vert\downarrow_1\rangle)$, so it is an eigenstate of linear combination of $\vec{\sigma}_1. \vec{\sigma}_3 $ and $ \vec{ \sigma}_2. \vec{\sigma}_3$. As $a$ and $b$ are connected to different local conserved quantities, due to evolution under Eq. (\ref{parental1}), such states do not couple with each other, and the dynamics is confined in the particular $a$, and $b$ block respectively. Hence, for the large $N$ limit, if the initial population is in a particular degenerate block, the degeneracy remains preserved and our result implies that $\mathcal{S}$ saturates with $N$.
\subsubsection{Dynamics confined in a linear combination of the multiple $J$ blocks}
If the initial state is the linear combinations of different $J$ blocks, then every block contributes to the entropy, and such contributions are additive. We provide here the analytical calculation of $\mathcal{S}$ for such a particular choice of the initial state, which is given by,
\begin{eqnarray}
\rh \sto &=& \sum\limits_{i = 1}^{\frac{N}{2} + 1/ \frac{N+1}{2}} \frac{\rho_i}{x_i}\\
 \sum\limits_{i = 1}^{\frac{N}{2} + 1/ \frac{N+1}{2}}  x_i &=& 1 \nn
\end{eqnarray}
Here $\rho_i$ is the density matrix of the $J= i$ sector. For even number of atoms, the upper limit is $\frac{N}{2} + 1$ and for odd number, it is $\frac{N+1}{2}$. The presence of $x_i $ leads to the modification of  $\mathcal{S}(i)$ in each blocks, which is given by,
\begin{eqnarray}
\mathcal{S}^\prime(i)=   \left(\frac{\mathcal{S}(i)}{x_i}+ \frac{\ln(x_i)}{x_i}\right)
\end{eqnarray}
For a large $N$ limit, the number of blocks $\approx \frac{N}{2}$. Here, we choose the initial condition to be equally populated with all the $J$ blocks to find the maximum growth of $\mathcal{S}$ as a function of $N$. In this case, $x_i = \frac{2}{N}$. We know that $\mathcal{S}(\frac{N}{2}) \geqslant \mathcal{S}(\frac{N}{2}- 1)$. To estimate the maximum growth, we further assume, $\mathcal{S}(i) =  \mathcal{S}(\frac{N}{2})$. Under this assumption,  the expression for the entropy at $\alpha =1$ is given by,
\begin{eqnarray}
\mathcal{S}_{\alpha=1} &= &  \sum\limits_{i = 1}^{\frac{N}{2} }  \mathcal{S}^\prime(i) \nn\\
&<& \mathcal{S}_{\alpha=1}\left(\frac{N}{2}\right) + \ln\left(\frac{N}{2}\right)
\end{eqnarray} 
As we show that for the large $N$ limit, $\mathcal{S}_{\alpha=1}(\frac{N}{2})$ is independent of $N$, our result implies that for the linear combination of all the $J$ blocks, a logarithmic growth of entropy $(\mathcal{S} )$ is found for increasing $N$. Such growth is negligible in comparison to the $\alpha < 1$ case. 

We provide two examples where the linear combination of different $J$ block is considered as an initial state. We choose, $\beta \omega_\circ = \ln(9)$. Considering two spin case, for $\vert \psi\rangle_{ini} = \vert 1 ,1 \rangle$, we find $\mathcal{S}= 0.382$. Next, for $\vert \psi\rangle_{ini} = \frac{1}{\sqrt{2}}(\vert 1 ,1 \rangle + \vert 0,0 \rangle)$, we get $\mathcal{S}= 0.884$. Similarly in case of three spins system, for $\vert \psi\rangle_{ini} = \vert \frac{3}{2}, \frac{3}{2} \rangle$, our results shows, $\mathcal{S} = 0.391$, and for $\vert \psi\rangle_{ini} = \vert \frac{1}{2}, -\frac{1}{2},a \rangle$, we get, $\mathcal{S} = 0.325$. Next, for $\vert \psi\rangle_{ini} = \frac{1}{\sqrt{2}}(\vert \frac{3}{2}, \frac{3}{2} \rangle + \vert \frac{1}{2}, -\frac{1}{2},a \rangle) $, we get $\mathcal{S} = 1.051$. Our result implies that for both cases, the maximum bound of entropy $(\mathcal{S}_{\rm total})$ is $ (\mathcal{S}_{\rm total}<\mathcal{S}(\text{principal block}) +  \ln(\text{no. of blocks}))$

Similarly, for $\alpha \to 1$, the system reaches a prethermal state whose growth of the entropy is negligible in comparison to the thermal state.

\section{Discussions}
\label{discussions}

We provide an example of a nearly integrable open quantum system, which shows the emergence of a prethermal plateau in the intermediate time scale. Our results have many similarities with the well-known properties of isolated many-body quantum systems \cite{Langen2016}. Following the works by Kollar \etal \cite{kollar2011}, a many-body system never thermalizes for evolving under an integrable Hamiltonian $H_\circ$, as such, the system relaxes to a nonthermal or GGE-described steady state. On the other hand, in the presence of perturbation $(H_\circ + g H_1)$,  prethermalization occurs in the inter-mediate timescale. Such a quasi-stationary prethermal state can be described by the same GGE-described state as mentioned above. Eventually, after a sufficiently long time, the system thermalizes due to the presence of the perturbation. Notably, $g=0$ acts as an integrable point. For the greater deviation of $g$ from the integrable point, the system quickly thermalizes, and the lifetime of the prethermal plateau becomes shorter.

Similarly, we consider a non-interacting many-body system coupled to a spatially correlated bosonic bath. The effect of the spatial correlation can be modeled using $\alpha(r/\xi)$. If  $r =0$, or $\xi \to \infty$, we note that $\alpha(r/\xi) =1$. On the other hand, when $r\gg \xi$, $\alpha(r/\xi) =0$. For the latter case, the whole system can be thought of as an identical copy of several single-atom systems coupled to its local environment. In such a case, the system relaxes to the Gibbs state, where $\beta$ is provided by the local environment. The interesting phenomenon occurs when $\alpha(r/\xi)=1$. The atoms become indistinguishable \emph{w.r.t} in the spatially correlated bosonic bath, which leads to several local conserved quantities in the system. For increasing the number of atoms, the number of the conserved quantities are ${N \choose 2}$. We show that such an integrable system relaxes to a generalized Gibbs state. We also study the property of such states. The GGE described states have initial value dependency; moreover, a pairwise entanglement is possible between the atoms. In the presence of a global symmetry operator, the density matrix can have a block-diagonal structure, where each block individually thermalizes. Due to such block-diagonalization, we found that the maximum growth of the entropy as a function of atom number is logarithmic, whereas such growth is linear for $\alpha(r/\xi) <1$.

We note that $\alpha(r/\xi) =1$ is an integrable point. Such a point appears when the atoms are very close $r_{mn} \approx 0$, or the bath correlation length becomes infinite. In real experiments, $\alpha(r/\xi) =1$ is not a physically achievable point. However, when the distance between the atoms is shorter than the bath correlation length, i.e., $\alpha(r/\xi) \to 1$, we find that such a system behaves like a nearly integrable system, and the dynamics involve multiple time scales. Our results imply that the system relaxes to a prethermal state in the intermediate time scale, where several quasi-conserved quantities constrain the dynamics. Such a prethermal state further relaxes to the thermal state after a sufficiently long time. The prethermal state is the steady state corresponding $\alpha =1$, and the thermal state is the steady state for $0 \leqslant \alpha(r/\xi) < 1$. For the value of $\alpha(r/\xi)$ is closer to the integrable point, a slowed relaxation to the thermal state is observed; as such, when $\alpha =1$, the prethermal state is the stationary state, which never relaxes. The key differences between the three types of open quantum systems (i.e., non-integrable, integrable, and nearly-integrable) are presented in the table \ref{table-1}, which appears for the different values of $\alpha(r/\xi)$. It is worth noting that a nearly integrable system is essentially a non-integrable system that displays integrability in the intermediate time scale. In terms of perturbation $(\delta)$, $\alpha(r/\xi) =1 - \delta$ \cite{mccutcheon_long-lived_2009}. The role of $\delta$ in our case is analogous to $g$,  for the case of the isolated many-body system, as described above \citep{kollar2011}.

\begin{table}[htb]
\begin{center}
\caption{Comparison between three different cases}
\begin{tabular}{p{3.0cm}| p{3.75cm}|  p{3.75cm} | p{5cm}  }
\hline
 \hline
 & Non-integrable & Integrable  & Nearly-integrable\\
 \hline
Condition & $0 \leqslant \alpha(r/\xi) <1$. & $\alpha(r/\xi) =1$. &  $\alpha(r/\xi) \to 1$.\\
Zero eigenvalues of $(\mathcal{\hat{L}})$ & One  & More than one  & One and ADR is close to zero.\\
(Quasi)steady state & Gibbs steady state. & Generalized-Gibbs steady state. & Generalized-Gibbs quasi-steady state followed by the Gibbs steady state.\\
(Quasi)conserved quantity & Does not exist. & ${N \choose 2}$ conserved quantities. & ${N \choose 2}$ quasi-conserved quantities.\\
  \hline
 \hline
 \end{tabular}
\label{table-1}
\end{center}
\end{table}

Our result exhibits that, for two qubits in a spatially correlated bath, the number of zero eigenvalues is two at $\alpha(r/\xi) =1$. Among them, the first one signifies the trace preservation or the conservation of total probability. The other zero eigenvalue indicates the presence of another conserved quantity, which is given in Eq. (\ref{conserved-1}). Such conserved quantity exists due to the creation and preservation of two spin correlations in the system. With the increase in the number of spins, along with the two-spin correlations (which increase by ${n \choose 2}$), higher spin correlations are also present in such cases. Identifying such higher spin correlations is beyond the interest of the present manuscript.

We have also shown that, for some specific choice of the initial condition, entanglement persists for two qubits system at $\alpha(r/\xi) = 1$. This can be readily extended to the $N$-spin system, where for $\alpha(r/\xi)=1$, the system evolves to a frozen steady state. Such a state has wide applications in the quantum storage devices, i.e., quantum batteries and environment-mediated information transfer devices \cite{Tabesh2020, Farina2019,kamin2020,gundogan2012,zhang_tunable_2018,zhang_generation_2021}.   
Persistent entanglement requires the physical proximity of the spins, which are coupled with the spatially-correlated bath. In such cases, the other source of relaxation (i.e., dipolar relaxation) becomes more effective. The effect of dipolar relaxation on the entanglement storage device was recently studied by the same authors \cite{saha_effects_2022}. It was shown that the dipolar coupling plays a detrimental role in quantum storage.

The bath correlation length is proportional to the bath lattice spacing $w_k$ \cite{jeske2013}. Hence, by suitably manipulating $w_k$, one can achieve the condition $r \ll \xi$ at a finite temperature where the Markovian approximation still holds. On the other hand, one can increase the bath correlation length by lowering the temperature. However, for a very low temperature, the Markovian approximation may not be valid as there may not be a clear separation between the bath and system timescale. The non-Markovian QME should be employed in such cases for the theoretical analysis of such systems \cite{Breuer_2016}.

\section{Conclusion}
\label{conclusion}
In summary, we show that for a collection of non-interacting atoms coupled to a spatially correlated Bosonic bath, the integrability depends on the spatial correlation function of the bath. If the distance between the atoms is much larger than the bath correlation length, then such a system acts as an identical copy of a single spin ensemble coupled to its own environment. The steady-state configuration can be obtained by using the Gibbs distribution. Therefore, it is an example of non-integrable OQS, where entropy is extensive.

On the other hand, when the atoms are overlapped or in the presence of an environment with diverging correlation lengths, we find that there exists an extensive number of conserved quantities as the Liouvillian has more than one zero eigenvalue. Moreover, a pairwise entanglement between the atoms is possible. Instead of Gibbs distribution, in this case, such steady-state configuration is given by a generalized Gibbs ensemble. The growth of entropy in such cases is much slower than the thermal state. Our result implies that maximum growth is logarithmic with increasing the number of atoms.

In our case, such an integrable point is an asymptotic point. We know that in experiments, a slow relaxation occurs when the distance between the atoms is shorter than the bath correlation length. As such, the system reaches a prethermal plateau in the intermediate time scale, and after a sufficiently long time, it further thermalizes. We find that the prethermal state is a generalized Gibbs state, whereas the thermal state is a Gibbs state. In the end, we propose the use of such an entangled prethermal state as a quantum storage
solution.

\begin{acknowledgments}
The authors thank Pradeep Kumar Mohanty, Subhasis Sinha, Arnab Chakrabarti, and Arpan Chatterjee for their insightful comments
and helpful suggestions. We also want to express our gratitude to the reviewer for the constructive criticism and valuable comments, which helped to improve the manuscript. SS gratefully acknowledges the University Grants Commission (UGC)
of Govt. of India for a research fellowship (Student ID: MAY2018- 528071).

\end{acknowledgments}

\appendix
\section{Calculation of $B(\omega_\circ)$, and $A(\omega_\circ)$}
\label{appendix-1}
The detailed form of the second order contribution of $\hsl^I$ in the QME (Eq. (\ref{qme-1})) for a single spin system is given by,
\begin{eqnarray}
&& \int\limits_0^\infty d\tau \tr_L [\hsl^I(t), [ \hsl^I(t-\tau), \rh(t)\otimes \rl]]^{\rm sec}\nn\\
&=&  \sum\limits_m \int\limits_0^\infty d\tau\left(e^{-i(\omega_\circ - \omega_m)\tau}\tr_L[\sigma_+a_m, [ \sigma_-a^{\dagger}_m, \rh(t)\otimes \rl]] + e^{i(\omega_\circ - \omega_m)\tau)}\tr_L[\sigma_-a^{\dagger}_m, [ \sigma_+a_m, \rh(t)\otimes \rl]] \right)\nn
\end{eqnarray}
Using the results from the Cauchy's integral,
\begin{eqnarray}
\int\limits_0^\infty d\tau e^{-i \omega \tau} = \pi \delta(\omega) - i P\left(\frac{1}{\omega}\right)
\end{eqnarray}
Here $P$ denotes the Cauchy principal value. The results corresponding to the partial trace over bath operators are given by,
\begin{eqnarray}
\tr_L[a^\dagger_m a_m \rl] &=& N(\omega_m)\nn\\
\tr_L[ a_m a^\dagger_m \rl] &=& 1 + N(\omega_m)\nn\\
\text{here}, \, N(\omega_m) &=& \frac{1}{e^{\beta \omega_m}-1}
\end{eqnarray}
In the continuum limit of the bath,  $\sum_m $ is replaced with $\int d^3 k/(2\pi)^3$. So,
\begin{eqnarray}
\sum\limits_m \to \frac{1}{(2\pi)^3 v^3} \int\limits_0^\infty 4\pi \omega_m^2 d\omega_m
\end{eqnarray}
Here, $\omega_m = v k_m $. For photon bath $v=c$. Using the above results, the form of the QME is given by,
\begin{eqnarray}
\frac{d \rh}{dt} &=&    \gamma_\circ (1 + N(\omega_\circ))(2\sigma_-\rh \sigma_+- \{\sigma_+\sigma_-, \rh\}) + \gamma_\circ N(\omega_\circ) (2\sigma_+\rh \sigma_-- \{\sigma_-\sigma_+, \rh\})
\label{qme-a-1}
\end{eqnarray}
This is known as the Lindblad master equation. The quantum dynamical map for the Lindblad equation holds complete positivity and trace preservation (CPTP).
Here, $\gamma_\circ = \frac{\wsl^2 \omega_\circ^2}{v^3}$. The effect of the imaginary terms is neglected, as they have provided a negligible shift to the Zeeman frequency $\omega_\circ$. We further use the constants $B(\omega_\circ),\, A(\omega_\circ)$ as
\begin{eqnarray}
B(\omega_\circ) &=& \gamma_\circ (1 + N(\omega_\circ)), \\
A(\omega_\circ) &=& \gamma_\circ N(\omega_\circ).
\end{eqnarray}
 
 \section{Derivation of the inhomogeneous dynamical equation from the QME }
\label{appendix-2}
 For a single spin system, we define $\rh $ as,
\begin{equation}
\rh = \begin{bmatrix}
 \rho_{11}  \quad & \rho_{12} \\  \rho_{21}  \quad &    \rho_{22} \end{bmatrix}  
\end{equation}
The dynamical equation (Eq. (\ref{qme-a-1})) in terms of the elements of $\rh$ is given by,
\begin{equation} 
\frac{d}{dt}\left[\begin{array}{c} \rho_{11} \\ \rho_{12}\\ \rho_{21} \\ \rho_{22}\end{array} \right] = \begin{bmatrix}
 -2B(\omega_\circ)  & 0 & 0 & 2A(\omega_\circ)\\ 0  &    -(A(\omega_\circ)+B(\omega_\circ)) & 0 &0\\ 0  &   0& -(A(\omega_\circ)+B(\omega_\circ))   &0 \\ 2B(\omega_\circ)  & 0 & 0 & -2A(\omega_\circ)\end{bmatrix}  \left[\begin{array}{c} \rho_{11} \\ \rho_{12}\\ \rho_{21} \\ \rho_{22}\end{array} \right]
\end{equation}
The above representation is known as the Liouville space representation. It is a homogeneous equation.  

From the above equation, we get,
\begin{eqnarray}
\frac{d}{dt} (\rho_{11} + \rho_{22})=0
\end{eqnarray}
Using the above constraint, the equation of $\rho_{11}$ is given by,
\begin{eqnarray}
\frac{d}{dt} \rho_{11}= -2(A(\omega_\circ)+B(\omega_\circ))\rho_{11} + 2A
 \end{eqnarray}

It is an inhomogeneous equation.
The trace preservation property transforms a homogeneous equation into an inhomogeneous one. The decomposition of $\rh$ in terms of Pauli matrices is given by,
\begin{eqnarray}
\rh(t) = \mathds{1}/2 + \sum\limits_{i=x,y,}^z\frac{M_i(t)}{2} \sigma_i,  
\end{eqnarray}
The presence of the identity matrix leads to the inhomogeneity in the equation. In terms of observables $(M_i)$ the dynamical equation is given by,
\begin{eqnarray}
\dot{M}_x &=& -R_1 M_x \nn\\
\dot{M}_y &=& -R_1 M_y \nn\\
\dot{M}_z &=& -2R_1 M_z  + 2M_\circ R_1\nn 
\end{eqnarray}
 Here we define, $R_1(\omega_\circ) = A(\omega_\circ)+B(\omega_\circ)$. In terms of $\beta$, we get $R_1(\omega_\circ) = \gamma_\circ(1 + 2N(\omega_\circ))= \gamma_\circ \coth (\beta \omega_\circ/2)$. $R_1(\omega_\circ)$ is called the relaxation rate and $M_\circ$ is the equilibrium magnetization.
 
 \section{Dynamical equation of the two spins case in terms of the observables. }
 \label{appendix-3}
 We note that for symmetric initial condition, the contribution of the anti-symmetric observables in the steady state is negligible. Using Eq. (\ref{2-spin-eq}), the dynamical equation of the nine symmetric observables is given by,
 \begin{eqnarray}
\frac{d}{dt} M_x &=& -R_1 M_x - 2\alpha M_\circ R_1 M_{xz}\nn\\
\frac{d}{dt} M_y &=& -R_1 M_y - 2\alpha M_\circ  R_1 M_{yz}\nn\\
\frac{d}{dt} M_z &=& -2 R_1 M_z + 2 M_\circ R_1 + 4\alpha R_1 (M_{xx}+M_{yy})\nn\\
\frac{d}{dt} M_{xx} &=& -2R_1 M_{xx} + 2 \alpha R_1 M_{zz} -  \frac{\alpha M_\circ R_1}{2} M_z\nn\\
\frac{d}{dt} M_{yy} &=& -2R_1 M_{yy} + 2 \alpha R_1 M_{zz} -  \frac{\alpha M_\circ R_1}{2} M_z\nn\\
\frac{d}{dt} M_{zz} &=& -4R_1 M_{zz} + M_\circ R_1 M_z + 2\alpha R_1 M_{xx} + 2\alpha R_1 M_{yy}\nn\\
\frac{d}{dt} M_{xz} &=& -(3 R_1 + 2 \alpha R_1) M_{xz} + \left(\frac{\alpha}{2} + 1 \right)M_\circ R_1 M_x \nn\\
\frac{d}{dt} M_{yz} &=& -(3 R_1 + 2 \alpha R_1) M_{yz} + \left(\frac{\alpha}{2} + 1 \right)M_\circ R_1 M_y\nn\\
\frac{d}{dt} M_{xy} &=& -2R_1 M_{xy}
\label{all-eq}
\end{eqnarray} 
The trace preservation leads to inhomogeneity in the dynamical equation of $M_z$. At the steady state, only $M_z$, $M_{zz}$, $M_{xx}$, and $M_{yy}$ will contribute.
 
 \section{Dynamical equation of the three spins case in terms of the relevant observables. }
  \label{appendix-4}
 For the three spins case, the number of observables is 63. For simplicity, we only present the dynamical equations of the relevant observables which contribute to the steady state.
 \begin{eqnarray}
\frac{d}{dt} M_z &=& -2R_1 M_z + 3 M_\circ R_1 + 4 \alpha M_\circ R_1 M_c\nn\\
\frac{d}{dt} M_c &=& -2M_c R_1 + 4 \alpha R_1 M_{zz} -4\alpha M_\circ R_1 M_{cz}-2 \alpha M_\circ R_1 M_z\nn\\
\frac{d}{dt} M_{zz} &=& -4 R_1 M_{zz}+2 M_\circ R_1 M_z - 2 \alpha R_1 M_c + 4 \alpha M_\circ R_1 M_{cz}\nn\\
\frac{d}{dt} M_{cz} &=& - 4 (1+ \alpha) R_1  M_{cz} + (1+ \alpha) M_\circ R_1 M_c + 12 R_1 \alpha M_{zzz} - 2\alpha M_\circ R_1 M_{zz}\nn\\
\frac{d}{dt} M_{zzz}&=& -6R_1 M_{zzz}+ 2\alpha R_1 M_{cz} + M_\circ R_1M_{zz} 
\end{eqnarray}
Here, $M_z = \tr[(\sum_{i=1}^3 I_z^i) \rh]$, $M_c = \tr[(\sum_{i,j=1}^3 (I_x^i\otimes I_x^j + I_y^i\otimes I_y^j)) \rh]$, $M_{zz} = \tr[(\sum_{i,j=1}^3 I_z^i\otimes I_z^j ) \rh]$, $M_{cz} = \tr[(\sum_{i,j,k=1}^3 (I_x^i\otimes I_x^j + I_y^i\otimes I_y^j) \otimes I_z^k)\rh]$, and $M_{zz} = \tr[(\sum_{i,j,k=1}^3 I_z^i\otimes I_z^j \otimes I_z^k) \rh]$. Here, $i \neq j \neq k$.
To find the steady-state solution, we need to solve the five coupled differential equations.

 \section{Comparison between the Zeeman basis result and the collective basis result }
  \label{appendix-5}
 We note that the dynamical equation derived in the Zeeman basis (Eq. (\ref{3eq})) and in the collective basis (Eq. (\ref{nspin-sol})) are coming from the same parent equation (Eq. (\ref{parental1})), hence they must provide the same solution at the steady state. We verify the above statement for the two choices of the initial states. The dynamical equation in the collective basis is given below.
 
For $N=2$, we get $J=1$. So, two different blocks exist, $J=0,\,1$. For the $J=0$ block, the dynamical equation is,
\begin{eqnarray}
\frac{d P_0}{dt} &=& 0
\end{eqnarray}
So $J=0$ block acts as a decoherence-free subspace. For the $J=1$ block, the dynamical equation is,
\begin{eqnarray}
\frac{d P_1}{dt} &=& -2(B(\omega_\circ) P_1 - A(\omega_\circ) P_{0})\nn\\
\frac{d P_0}{dt} &=& -2(A(\omega_\circ)P_0 - B(\omega_\circ) P_{1}) - 2(B(\omega_\circ) P_0 - A(\omega_\circ) P_{-1})\nn\\
\frac{d P_{-1}}{dt} &=& -2(A(\omega_\circ)P_{-1} - B(\omega_\circ) P_{0})  \nn
\end{eqnarray}
We note that, $\dot{P}_1 + \dot{P}_0 + \dot{P}_{-1} = 0$. Such constraint leads to inhomogeneity in the dynamical equation. The steady state solution requires, $P_i/P_{i-1} = \exp(\beta \omega_\circ)$, which signifies every block separately thermalizes. 

The dynamical equation in the Zeeman basis is given below,
 \begin{equation}
\left[\begin{array}{c} \dot{M}_z \\ \dot{M}_{zz}\\ \dot{M}_{c} \end{array} \right] = \begin{bmatrix}
-2R_1  & 0 & 4M_{\circ}\alpha R_1 \\   M_{\circ}R_1& -4R_1  & 2 \alpha R_1  \\ 
-M_{\circ} \alpha R_1 &\,\,\, 4 \alpha R_1  & -2R_1  \end{bmatrix}  \left[\begin{array}{c} M_z  \\
M_{zz}\\ M_c \end{array}\right] + \left[\begin{array}{c} 2M_{\circ} R_1 \\  0\\ 0 \end{array}\right]  
\end{equation} 
 We note that, for $\alpha =1$, $\dot{M}_{zz}+ \dot{M}_{c} = 0$. Two particular cases are studied below,
\begin{itemize}
\item If the initial state is singlet state, $\vert J=0, M=0 \rangle$, then the solutions for the two equations are given by,
$P_0 \sts= 1$ for $J=0$ and $P_1 \sts=P_{-1}\sts=P_0\sts=0$ for $J=1$. On the other hand, in the Zeeman basis, $F = -\frac{3}{4}$, so $M_z =0$, $M_c = -\frac{1}{2}$, $M_{zz} = -\frac{1}{4}$. Hence, $\rh\sts = \frac{\mathds{I}}{4} - I_{z}\otimes I_z - (I_x\otimes I_x + I_y \otimes I_{y})$. In both cases, the final form of $\rh\sts$ is the same.
\item If the initial state is $\vert \uparrow \uparrow \rangle$, then the solutions for the two equations are given by,
$P_0 = 0$ for $J=0$ and $P_1= \frac{e^{-\beta \omega_\circ}}{e^{-\beta \omega_\circ} + 1 + e^{\beta \omega_\circ}}$, $ P_{0}=\frac{1}{e^{-\beta \omega_\circ} + 1 + e^{\beta \omega_\circ}}$, and $P_{-1}= \frac{e^{\beta \omega_\circ}}{e^{-\beta \omega_\circ} + 1 + e^{\beta \omega_\circ}}$  for $J=1$, whereas, in the Zeeman basis, $F = \frac{1}{4}$, so, $M_z = \frac{4 M_\circ}{M_\circ^2 + 3}$, $M_c = \frac{1- M_\circ^2}{2 (M_\circ^2 + 3)}$, and $M_{zz} = F-M_c$. Using this three operator, the eigenvalues of $\rh$ at the steady state are $\{0, \frac{e^{-\beta \omega_\circ}}{e^{-\beta \omega_\circ} + 1 + e^{\beta \omega_\circ}}, \frac{1}{e^{-\beta \omega_\circ} + 1 + e^{\beta \omega_\circ}},  \frac{e^{\beta \omega_\circ}}{e^{-\beta \omega_\circ} + 1 + e^{\beta \omega_\circ}}\}$ . 

\end{itemize}
 Hence, both equations provide the same solutions at the steady state.

\bibliographystyle{apsrev4-1}
\bibliography{references}
\end{document}